# SPP location with spherical ray tracing by refractive index

Hui Qian    Xiaosan Zhu    Dongliang Liu

Institute of Geology, Chinese Academy of Geological Sciences

**Abstract:** Atmospheric layer structure is a primary factor affecting the precision of single-point satellite positioning. The assumption of electromagnetic wave rectilinear propagation hinders the accurate implementation of ionospheric and tropospheric corrections, whereas curvilinear positioning methods fully account for ray deflection. This study aims to derive partial derivative formulas for theoretical travel time with respect to latitude, longitude, elevation, and velocity models by formulating electromagnetic wave travel time equations under a coordinate-based one-dimensional layered velocity model. Subsequently, a linearized LSQR method is employed to invert station coordinates, receiver clock biases, and electromagnetic wave velocities at the bottom of the ionosphere and troposphere using over six sets of observations. This replaces conventional ionospheric/tropospheric pseudorange corrections in single-point positioning, establishing a novel spherical coordinate refraction travel time positioning method. The classical straight-line pseudorange positioning is reformulated into a time-of-flight positioning approach, and the positioning accuracy differences between straight-line and spherical coordinate refraction travel time methods are compared. By integrating classical ionospheric and tropospheric models to construct corresponding refractive index models and combining them with curvilinear ray tracing methods, the inherent theoretical limitations of positioning can be effectively mitigated.

## Introduction

Global Navigation Satellite System (GNSS) positioning algorithms have traditionally operated under the fundamental assumption that radio signals propagate in straight - line paths between satellites and receivers. This assumption, while simplifying the underlying mathematical models, fails to account for the significant atmospheric effects that can substantially distort signal propagation. Atmospheric refractivity, which varies spatially due to factors such as temperature, pressure, and humidity gradients, causes radio signals to deviate from their straight - line trajectories through refraction. This refraction phenomenon introduces systematic errors in position estimates, particularly in scenarios where high - precision positioning is crucial.

Conventional approaches to mitigate these errors have relied on empirical models. For example, the Klobuchar model is widely used for ionospheric delay corrections in single - frequency GNSS receivers (Klobuchar, 1987). The NeQuick model, on the other hand, is a more sophisticated ionospheric correction model that takes into account solar activity and other parameters (Nava et al., 2008). However, these empirical models have inherent limitations. Their accuracy is often compromised in dynamic environments, such as those with rapidly changing atmospheric conditions, or during extreme weather events. In such cases, the fixed empirical relationships used in these models may not accurately represent the actual atmospheric state, leading to residual errors in position calculations.In recent years, curve - based ray tracing (CRT) algorithms have emerged as a promising solution to enhance GNSS positioning accuracy, especially in challenging environments like urban canyons and indoor spaces (Misra and Enge, 2006). Unlike traditional line - of - sight (LOS) models that assume straight - line signal propagation, CRT algorithms model signal paths as curves. This allows them to account for various wave phenomena such as diffraction, reflection, and scattering. By doing so, CRT algorithms can mitigate multipath errors, which occur when signals reach the receiver through multiple paths due to reflections from nearby objects, and improve positioning reliability in

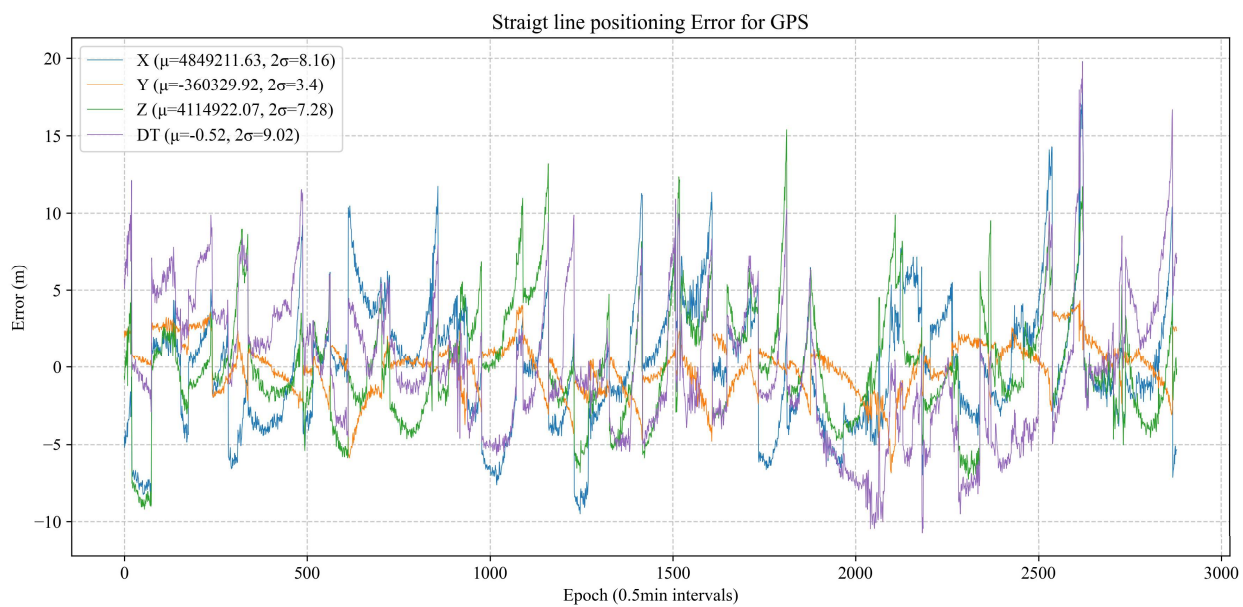

Fig.1　Error of GPS observation positioning with straight line approximation

non - line - of - sight (NLOS) conditions.Atmospheric refractivity gradients, particularly in the troposphere and ionosphere, are the primary causes of signal bending. For instance, the tropospheric zenith delay, which is the delay in signal propagation caused by the troposphere when the satellite is at the zenith, can exceed 2.5 meters under certain atmospheric conditions (Boehm et al., 2006). In multi - GNSS constellations, where low - elevation satellites play a critical role in urban canyon navigation, these errors can compound. The ray curvature in the troposphere, which is a result of the vertical refractivity gradient, enables dynamic path adjustments. Despite their advantages, CRT algorithms still face several challenges. In dynamic environments, such as when the receiver is in a moving vehicle, the rapidly changing atmospheric conditions and the motion of the receiver make it difficult to accurately model the signal paths.

Additionally, CRT algorithms require high - resolution digital elevation models (DEMs) to accurately represent the terrain and surrounding environment. In deep urban canyons, atmospheric delays and signal blockage due to tall buildings further complicate the signal propagation and pose unresolved issues for CRT algorithms.Curvilinear tracing, by its very nature, is well - suited to accommodate wave phenomena like diffraction and reflection. By integrating Fresnel - Kirchhoff diffraction integrals with ray tracing, algorithms can more effectively resolve NLOS signals and mitigate multipath biases (Ishimaru, 1997). In contrast, linear positioning algorithms, although computationally efficient, are inherently limited by their oversimplification of atmospheric physics. They fail to capture the complex interactions between the radio signals and the atmosphere, which are essential for achieving high - precision positioning.Curvilinear ray tracing, which is grounded in the first principles of wave propagation and refractivity profiling, offers a potential path to sub - meter accuracy even in challenging environments. This paper advocates for a paradigm shift from the traditional linear ray assumption to curvilinear ray tracing in GNSS positioning algorithms. By leveraging atmospheric refractivity profiles to model the true signal trajectories, it is hoped that a new generation of GNSS positioning algorithms can be developed that can overcome the limitations of current approaches and provide more accurate and reliable positioning information in a wide range of applications.

**Methodology**

Assuming the refractive index from satellite to top of atmosphere is 1, with electromagnetic wave velocity as light speed $c$. The atmosphere can be divided into multiple layers. For a specific layer with bottom/top velocities $v_1, v_2$ and radii $r_1, r_2$, the wave velocity $v$ within this layer follows the spherical coordinate formula centered at Earth: $v = ar^{1-b}$, where

$$a = \frac{-\ln(r_2)\ln(v_1) + \ln(r_1)\ln(v_2)}{\Delta r} \quad b = 1 - \frac{\Delta v}{\Delta r}$$

and $\Delta r = \ln\left(\frac{r_1}{r_2}\right) \Delta v = \ln\left(\frac{v_1}{v_2}\right)$. For incident electromagnetic wave with ray parameter $p$, the travel time through this layer is

$$\Delta t = \frac{-u_1 + u_2}{b}$$

where

$$u_1 = \sqrt{-p^2 + \frac{r_1^2}{v_1^2}} \quad u_2 = \sqrt{-p^2 + \frac{r_2^2}{v_2^2}}$$

, The horizontal arc traversed by the wave in this layer is:

$$\Delta w = \frac{-\text{ArcTan}\left(\frac{u_1}{p}\right) + \text{ArcTan}\left(\frac{u_2}{p}\right)}{b}$$

For ray tracing, we require the derivative

$$\frac{\partial w}{\partial p} = \frac{1}{bu_1} - \frac{1}{bu_2}$$

to determine $p$ under known horizontal distance constraints. Since $p$ remains constant across layers for a single ray, horizontal distance equals the sum of all layer-specific horizontal arcs. The ray parameter $p$ can be solved via damped least squares optimization.

To invert near-surface refractive indices and station heights, we need derivatives of time and distance with respect to velocity $v$ and spherical radius $r$:

$$\frac{\partial w}{\partial r_1} = \frac{b\Delta r^2 p + \Delta v \Delta w u_1}{b^2 \Delta r^2 r_1 u_1} \quad \frac{\partial w}{\partial r_2} = \frac{b\Delta r^2 p + \Delta v \Delta w u_2}{b^2 \Delta r^2 r_2 u_2}$$

$$\frac{\partial t}{\partial r_1} = \frac{-b\Delta r^2 r_1^2 - \Delta v v_1^2 u_1 - u_1 + u_2}{b^2 \Delta r^2 r_1 v_1^2 u_1} \quad \frac{\partial t}{\partial r_2} = \frac{b\Delta r^2 r_2^2 + \Delta v v_2^2 u_2 - u_1 + u_2}{b^2 \Delta r^2 r_2 v_2^2 u_2}$$

$$\frac{\partial w}{\partial v_1} = \frac{b\Delta r p + \Delta w u_1}{b^2 \Delta r v_1 u_1} \quad \frac{\partial w}{\partial v_2} = \frac{-b\Delta r p - \Delta w u_2}{b^2 \Delta r v_2 u_2}$$

$$\frac{\partial t}{\partial v_1} = \frac{b\Delta r r_1^2 + \Delta t v_1^2 u_1}{b^2 \Delta r v_1^3 u_1} \quad \frac{\partial t}{\partial v_2} = \frac{-b\Delta r r_2^2 - \Delta t v_2^2 u_2}{b^2 \Delta r v_2^3 u_2}$$

Since horizontal distance equals the sum of layer-specific horizontal arcs, a cross-summation process applies to upper/lower layers for derivatives of total distance and time with respect to refractive index altitude $z$ and velocity $v$. For ray parameters, this becomes a cumulative summation. When interpolating near-surface refractive indices using exponential functions:

$$v_0 = e^{a+1-b\ (z_0)} \qquad \frac{\partial v_0}{\partial z_0} = \frac{\Delta v v_0}{\Delta r z_0}$$

$$\frac{\partial v_0}{\partial r_1} = \frac{\Delta v \ln\left(\frac{r_2}{z_0}\right) v_0}{\Delta r^2 r_1} \qquad \frac{\partial v_0}{\partial r_2} = -\frac{\Delta v \ln\left(\frac{r_1}{z_0}\right) v_0}{\Delta r^2 r_2}$$

$$\frac{\partial v_0}{\partial v_1} = -\frac{\ln\left(\frac{r_2}{z_0}\right) v_0}{\Delta r r_1} \qquad \frac{\partial v_0}{\partial v_2} = \frac{\ln\left(\frac{r_1}{z_0}\right) v_0}{\Delta r r_2}$$

For interpolated layers $z_1, z_2$ :

$$\frac{\partial w}{\partial z_1} = \frac{\partial w}{\partial r_1} + \frac{\partial w}{\partial v_0}\frac{\partial v_0}{\partial z_0} \quad \frac{\partial t}{\partial z_1} = \frac{\partial t}{\partial r_1} + \frac{\partial t}{\partial v_0}\frac{\partial v_0}{\partial z_0} \quad \frac{\partial w}{\partial z_2} = \frac{\partial w}{\partial r_2} + \frac{\partial w}{\partial v_0}\frac{\partial v_0}{\partial z_0} \quad \frac{\partial t}{\partial z_2} = \frac{\partial t}{\partial r_2} + \frac{\partial t}{\partial v_0}\frac{\partial v_0}{\partial z_0}$$

Additional compensation terms are required for interpolated layers:

$$\frac{\partial w}{\partial z} += \frac{\partial w}{\partial v_0}\frac{\partial v_0}{\partial r_1}\frac{\partial t}{\partial z} += \frac{\partial t}{\partial v_0}\frac{\partial v_0}{\partial r_1}\frac{\partial w}{\partial v} += \frac{\partial w}{\partial v_0}\frac{\partial v_0}{\partial v_1}\frac{\partial t}{\partial v} += \frac{\partial t}{\partial v_0}\frac{\partial v_0}{\partial v_1}$$

The ray parameter p can be obtained by inversion of the summation that approximates the actual distance to dw, completing ray tracing. Here, damped least squares or other optimization methods that utilize known derivative information can be employed. After inversion, when calculating the derivatives of the traced ray paths for each layer and the top/bottom heights, it is crucial to subtract the derivative of the distance multiplied by the factor $\frac{\partial t}{\partial w}$, as the distance information (used in the inversion) has been coupled. Since $\frac{\partial t}{\partial w}$ for each layer is equal to p, the overall summed $\frac{\partial t}{\partial w}$ remains equal to p. Therefore, it follows that:

$$\frac{\partial t}{\partial z} -= p\frac{\partial w}{\partial z} \quad \frac{\partial t}{\partial v} -= p\frac{\partial w}{\partial v} \quad \frac{\partial t}{\partial z_1} -= p\frac{\partial w}{\partial z_1} \quad \frac{\partial t}{\partial z_2} -= p\frac{\partial w}{\partial z_2}$$

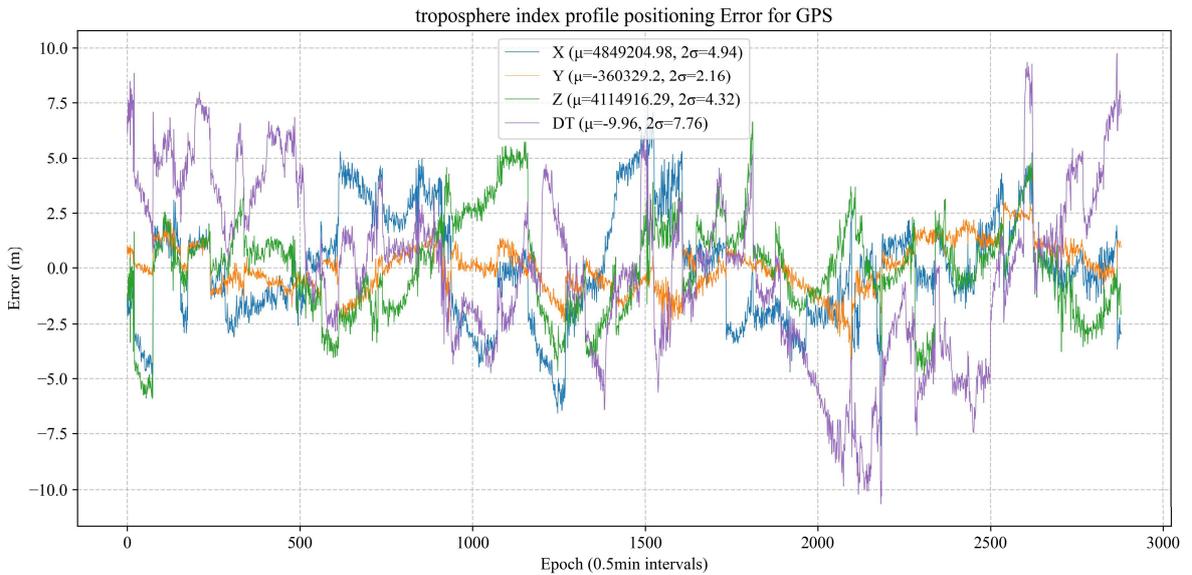

Fig.2　Error of GPS observation positioning with troposphere refraction index

Using spherical arc length
$$w = ArcCos(cos(j_1 - j_2)cos(w_1)cos(w_2) + sin(w_1)sin(w_2))$$
we derive:

$$\frac{\partial w}{\partial w_1} = -\frac{-cos(j_1 - j_2)cos(w_2)sin(w_1) + cos(w_1)sin(w_2)}{sin(w)}$$

$$\frac{\partial w}{\partial j_1} = \frac{cos(w_1)cos(w_2)sin(j_1 - j_2)}{sin(w)}$$

$$\frac{\partial w}{\partial w_2} = -\frac{cos(w_2)sin(w_1) - cos(j_1 - j_2)cos(w_1)sin(w_2)}{sin(w)}$$

$$\frac{\partial w}{\partial j_2} = -\frac{cos(w_1)cos(w_2)sin(j_1 - j_2)}{sin(w)}$$

Time derivatives for inversion parameters are:
$$\frac{\partial t}{\partial w_1} = p\frac{\partial w}{\partial w_1} \quad \frac{\partial t}{\partial j_1} = p\frac{\partial w}{\partial j_1} \quad \frac{\partial t}{\partial w_2} = p\frac{\partial w}{\partial w_2} \quad \frac{\partial t}{\partial j_2} = p\frac{\partial w}{\partial j_2}$$

Combining with known station positions and refractive index profiles can be inverted from time differences.

In practice, near-surface refractive indices can be inverted directly or estimated via ionospheric and tropospheric models. Station positions are primarily inverted through curved ray tracing, with accuracy dependent on satellite quantity and data quality.

**Application**

For GPS observation P1 data acquired on UTC July 20, 2006, we initially performed straight-line positioning without tropospheric and ionospheric corrections. This resulted in a position error of 8 meters and a pseudorange error of 9 meters(Fig.1).When adopting a tropospheric model with a refractive index profile defined by 293 at 10 km below ground level, 269.1694 at ground level, and 0 at 12 km altitude, the positioning accuracy was significantly improved. The position error was reduced to 5 meters, and the pseudorange error decreased to 8 meters, demonstrating clear enhancements compared to the uncorrected straight-line positioning(Fig.2)

Upon introducing the ionospheric refractive index model using IRI-2020 to obtain electron density $N_e$, we derived the corresponding altitude-resolved refractive index profile for fixed latitude/longitude at the observation time, as shown in following table

| 10 | 0 | -12 | -20 | -30 | -50 | -80 | -100 | -150 | -200 | -300 | -450 | -600 | -800 | -1000 |
|---|---|---|---|---|---|---|---|---|---|---|---|---|---|---|
| 293 | 269.17 | 0 | 0 | 0 | 0 | 0.0084 | 0.0285 | 0.0153 | 0.4410 | 4.8929 | 2.0460 | 0.7356 | 0.2788 | 0 |

The results indicate that ionospheric refractive indices exhibit minimal values, exerting limited influence on positioning accuracy. The errors remained consistent with those after tropospheric corrections alone(Fig.3).

Simultaneous application of conventional tropospheric and ionospheric corrections yielded further reduced errors: a position error of 4.6 meters and pseudorange error of 8 meters. This suggests residual inaccuracies in the refractive index profile(Fig.4). More precise determination of tropospheric refractive indices requires inversion algorithms, as mentioned in the methodology section. While such inversions are feasible, they demand higher satellite numbers and data quality, which were not pursued in this study. Nonetheless, our findings confirm that the proposed refractive index profiling algorithm effectively realizes combined tropospheric and ionospheric corrections. Further refinements are warranted to address remaining uncertainties in the profile modeling.

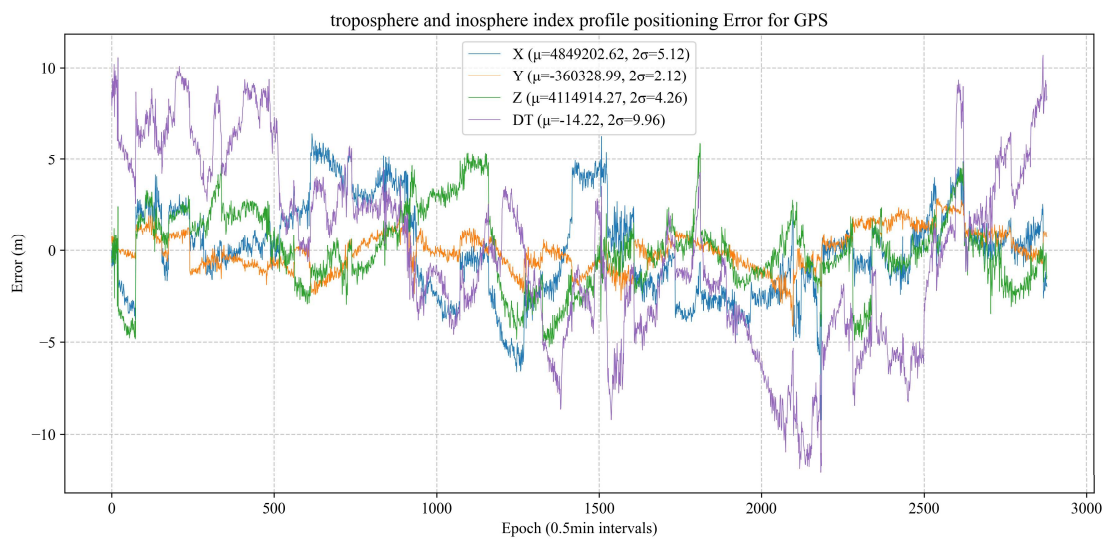

Fig.3　Error of GPS observation positioning with troposphere and inosphere refraction index

**Disccussion**

In the field of Global Positioning System (GPS) positioning, the accuracy of measurements can be significantly influenced by atmospheric conditions, notably the troposphere and ionosphere. This discussion synthesizes the effects of tropospheric and ionospheric corrections on GPS observations. The analysis compares direct positioning results with and without atmospheric corrections, revealing insights into the importance of employing such models for improving positioning accuracy. Without applying corrections for either the troposphere or ionosphere, the positioning error recorded was approximately 8 meters, along with a pseudorange error of about 9 meters. These values illustrate the inherent inaccuracies in GPS positioning due to atmospheric interference, particularly caused by the variations in atmospheric refractivity. Tropospheric delay contributes significantly to the overall positioning error. The tropospheric delay is primarily affected by water vapor, which is not uniformly distributed and varies with atmospheric conditions, thus necessitating accurate modeling and correction to improve positioning outcomes (Ibrahim & El-Rabbany, 2011). Upon integrating a tropospheric correction model, a notable improvement was observed: the location error decreased to 5 meters with the time pseudorange error reducing to 8 meters (Kedar et al., 2003). Research corroborates this improvement, indicating that rigorous correction for tropospheric effects can enhance GPS height determination significantly (Wang et al., 2008). The study emphasizes the critical need for accurate meteorological data to refine tropospheric corrections, validating the utility and effectiveness of such approaches in GPS applications. The introduction of an ionospheric correction using the IRI-2020 model showcased implications for positioning accuracy. The effective implementation of correction models can mitigate significant errors in positioning due to ionospheric delay (Hoque et al., 2018). Although the ionospheric delay is a crucial source of error, the impact of the corrections provided by the IRI-2020 model is variable, as there is some debate in the literature regarding its applicability in different conditions (Kedar et al., 2003). When applying a combined correction for both tropospheric and ionospheric effects, the results were further refined, yielding a positioning error of 4.6 meters and maintaining a time pseudorange error of 8 meters. This outcome underscores the effectiveness of dual-correction techniques in GPS positioning systems, suggesting that weather and atmospheric modeling can contribute to enhanced accuracy in spatial orientation (Singh & Reilly, 2006). In recognizing the layered complexities of atmospheric effects on GPS signals, it becomes essential to develop sophisticated algorithms that enable real-time adjustments and corrections based on extensive satellite and reference station data. Advanced inversion algorithms to refine the understanding and modeling of atmospheric refractivity profiles have been discussed (Wadge et al., 2002).

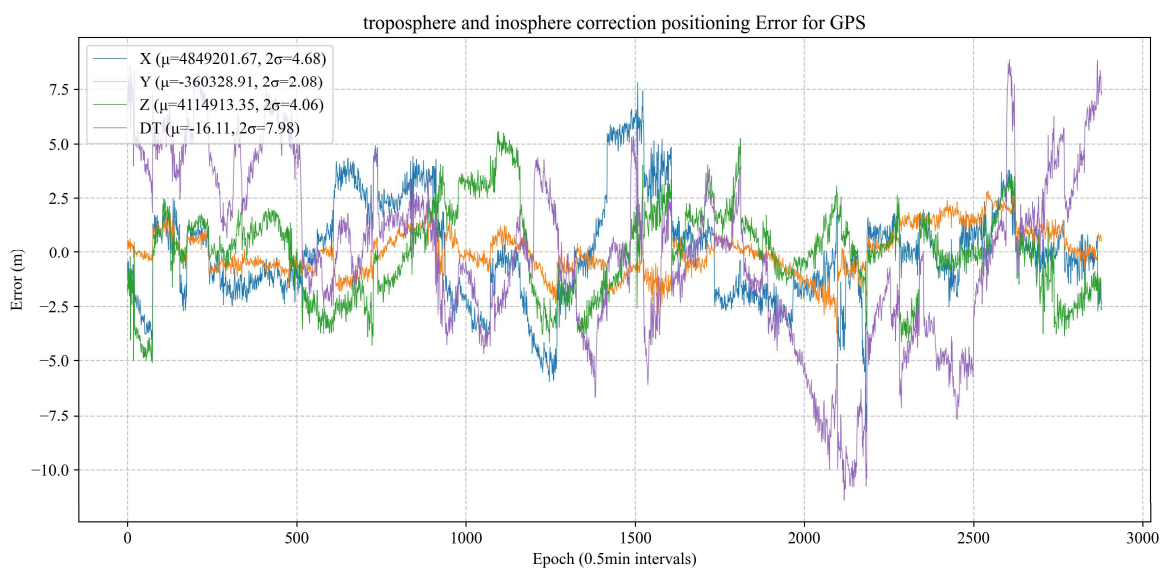

Fig.4　Error of GPS observation positioning with ordinary troposphere and inosphere correction

However, this approach requires substantial computational resources and high-quality input data, a challenge that must be addressed for practical implementation (Ibrahim & El-Rabbany, 2011). Furthermore, studies have indicated that enhanced positioning accuracy can also be attained through localized differential systems that integrate tropospheric corrections based on dynamic meteorological inputs (Zhou & Wu, 2014). These systems could significantly reduce the errors associated with single-frequency GPS systems, which are particularly susceptible to atmospheric influences (Zhang et al., 2021). Additionally, continuous monitoring of the ionosphere and utilization of real-time data streams can facilitate robust modeling that dynamically adjusts for seasonal and diurnal variations in atmospheric conditions. The use of advanced tools such as the Klobuchar model, which incorporates real-time ionospheric corrections based on GPS signals, has been substantiated as effective in reducing overall positioning errors (Li et al., 2005). In conclusion, the synthesis of atmospheric corrections, both tropospheric and ionospheric, reveals a marked improvement in GPS positioning accuracy. The reduced positioning error achieved underscores the necessity of employing robust meteorological models to enhance the performance of GPS systems, making them more reliable and efficient for various applications. Future research should focus on refining these models further and integrating real-time atmospheric data to continually enhance the accuracy of GPS positioning across diverse environments.

**Conclusion**

The methodologies employed in mitigating both tropospheric and ionospheric delays establish significant advancements in accuracy. The following points outline the availability and innovative aspects of the new method:
1. Dynamic Meteorological Adaptation: New approaches that utilize real-time meteorological data for tropospheric corrections reflect a trend toward adaptability. This development allows for corrections that are sensitive to immediate atmospheric conditions, further reducing errors attributable to variable atmospheric refractivity.
2. Computational Advances in Algorithm Development: Innovations in sophisticated algorithms for real-time atmospheric correction adjustments signify a forward-thinking approach in the field. Advanced inversion techniques, relying on extensive satellite data and local weather models, are being researched to improve the understanding of atmospheric impacts on GPS signals
3. Increased Resilience to Atmospheric Variability: Continuous monitoring and analysis of both tropospheric and ionospheric conditions through advanced techniques signify innovation in resilience against atmospheric interferences. Research indicates that sustained monitoring can lead to robust models that dynamically adjust for seasonal and diurnal atmospheric changes, ensuring GPS positioning remains reliable under varied conditions

**Author contribution**

Hui Qian: Conceptualization, Methodology, Software, Formal Analysis, Writing
Xiaosan Zhu: Data Curation, Writing - Review & Editing, Funding Acquisition
Dongliang Liu: Visualization, Supervision

**Open Research**

The compiled program together with IRI2020 model and test data are available at https://figshare.com/s/8447f8ac1b9307fd8fa7 or 10.6084/m9.figshare.29133758

**Acknowledgment**

This work is funding by the Ministry of Science and Technology of the People's Republic of China (No. 2019YFA0708601-02),and the Scientific Research Fund Project of BGRIMM Technology Group (No. JTKY202427822)